\documentclass{elsart}

\usepackage{graphicx}

\usepackage{amssymb}

\begin{document}

\begin{frontmatter}



\title{Studies of Proximity Focusing RICH \\ 
with an aerogel radiator using Flat-panel multi-anode PMTs (Hamamatsu H8500)}


\author[a]{T.~Matsumoto\corauthref{cor}}
\ead{matumot@bmail.kek.jp}
\author[b]{S.~Korpar}
\author[c]{I.~Adachi}
\author[b]{S.~Fratina}
\author[d]{T.~Iijima}
\author[e]{R.~Ishibashi}
\author[e]{H.~Kawai}
\author[b]{P.~Kri\v zan}
\author[f]{S.~Ogawa}
\author[b]{R.~Pestotnik}
\author[c]{S.~Saitoh}
\author[a]{T.~Seki}
\author[a]{T.~Sumiyoshi}
\author[c]{K.~Suzuki}
\author[e]{T.~Tabata}
\author[f]{Y.~Uchida}
\author[e]{Y.~Unno}

\address[a]{Tokyo Metropolitan University, Tokyo, Japan}
\address[b]{Jo\v{z}ef Stefan Institute, Ljubljana, Slovenia}
\address[c]{High Energy Accelerator Research Organization (KEK), Tsukuba, Japan}
\address[d]{Nagoya University, Nagoya, Japan}
\address[e]{Chiba University, Chiba, Japan}
\address[f]{Toho University, Funabashi, Japan}

\corauth[cor]{Corresponding author. Tel: +81-426-77-2500; fax: +81-426-77-2483}

\begin{abstract}
A proximity focusing ring imaging Cherenkov detector using aerogel as the radiator 
has been studied for an upgrade of the Belle detector at the KEK-B-factory. 
We constructed a prototype Cherenkov counter using a $4{\times}4$ array of 
64-channel flat-panel multi-anode PMTs (Hamamatsu H8500) with a large effective 
area. The aerogel samples were made with a new technique to obtain a 
higher transmission length at a high refractive index ($n=1.05$). Multi-channel 
PMTs are read-out with analog memory chips. The detector was tested at the 
KEK-PS $\pi2$ beam line in November, 2002. To evaluate systematically the 
performance of the detector, tests were carried out with various aerogel samples 
using pion beams with momenta between 0.5~GeV/$c$ and 4~GeV/$c$. The typical 
angular resolution was around 14~mrad, and the average number of 
detected photoelectrons was around 6. We expect that pions and kaons can be 
separated at a 4$\sigma$ level at $4~{\rm GeV}/c$.
\end{abstract}

\end{frontmatter}



\section{Introduction}

Silica aerogel is a unique material with a refractive index ($n$) in the range 
between gases and liquids or solids. Its refractive index can be easily controlled 
from $n=1.01$ to $1.06$. As a result, the refractive index of the aerogel can be 
chosen such that for a given momentum interval in the few GeV/$c$ region charged 
pions radiate Cherenkov photons, while kaons stay below the  Cherenkov 
radiation threshold \cite{AeroCounter}. In the Belle experiment at KEK \cite{Belle}, 
a threshold type Cherenkov detector (Belle-ACC) \cite{BelleACC} which uses aerogel 
as a radiator, is operated, providing at $3.5~{\rm GeV}/c$ a kaon identification 
efficiency of $88$\% with a pion missidentification probability of $8$\% 
\cite{BelleKID}. 

A new production method of hydrophobic aerogel with a high transmission length and 
$n$ in the interval between $1.01$ and $1.03$ was developed during the 
construction period of Belle-ACC \cite{BelleAerogel}. The improvement in 
quality allows the use of an aerogel radiator in a ring imaging Cherenkov 
counter (RICH) \cite{AeroRICH}. In the HERMES experiment at DESY, a RICH counter 
is used with a dual-radiator (aerogel and gas), and mirrors to focus the 
Cherenkov photons \cite{HERMES_RICH}. A similar detector is also designed for the 
LHCb experiment at CERN \cite{LHCB_RICH}.  

We are studying the feasibility of a RICH counter with an aerogel radiator for the 
Belle-ACC in the forward end-cap region \cite{BelleAeroRICH}. Since this part is 
now optimized for the pion/kaon separation needed for tagging of the $B$ flavor, 
and covering the momentum range below $2~{\rm GeV}/c$, separation at high-momentum 
region of around $4~{\rm GeV}/c$ is not adequate. This kinematic region is, however, 
very important for the studies of two-body decays such as $B \to \pi\pi$, $K\pi$. 
In order to achieve a $\pi$/$K$ separation for a wider momentum range, a ring 
imaging-type of detector is needed. Due to spatial restrictions, such a counter has 
to be of the proximity focusing type. To cover the identification in the lower 
momentum region (around $0.7~{\rm GeV}/c$) as well as in the region up to 
$4~{\rm GeV}/c$, the aerogel has to have a refractive index around $n = 1.05$. The 
first beam test of such a detector was carried out in 2001 at the KEK-PS $\pi2$ beam 
line \cite{1stBeamTest}. These tests used an array of multi-anode PMTs (Hamamatsu 
R5900-M16) for photo-detection. The detected number of photoelectrons was $2.7$ per 
ring for a $2~{\rm cm}$ thick aerogel tile with $n = 1.05$, and the Cherenkov 
angle resolution per photon was $10~{\rm mrad}$. These results were consistent 
with expectations. The number of detected photons was, however, rather low, 
partly because only $36$\% of the detector surface was covered by the photo-cathodes, 
and partly because the transmission length of the aerogel with $n = 1.05$ could not 
be made large enough. For the second beam test, we improved the aerogel 
transmission by optimizing the materials used in the production process. 
The active area fraction of the photon detector was increased by employing 
recently developed flat-panel PMTs, Hamamatsu H8500. Although this type of PMT is 
not immune to magnetic field, and therefore cannot be applied in the Belle 
spectrometer, we consider this device as an intermediate step in our 
development. The paper is organized as follows. We first present the experimental 
set-up with flat-panel PMTs, briefly review the improvement in aerogel 
production, describe the measurements, and finally discuss the results.

\section{The experiment set-up}

\subsection{Flat-panel PMT}

\begin{figure}[p]
  \begin{center}
    \includegraphics[keepaspectratio=true,width=120mm]{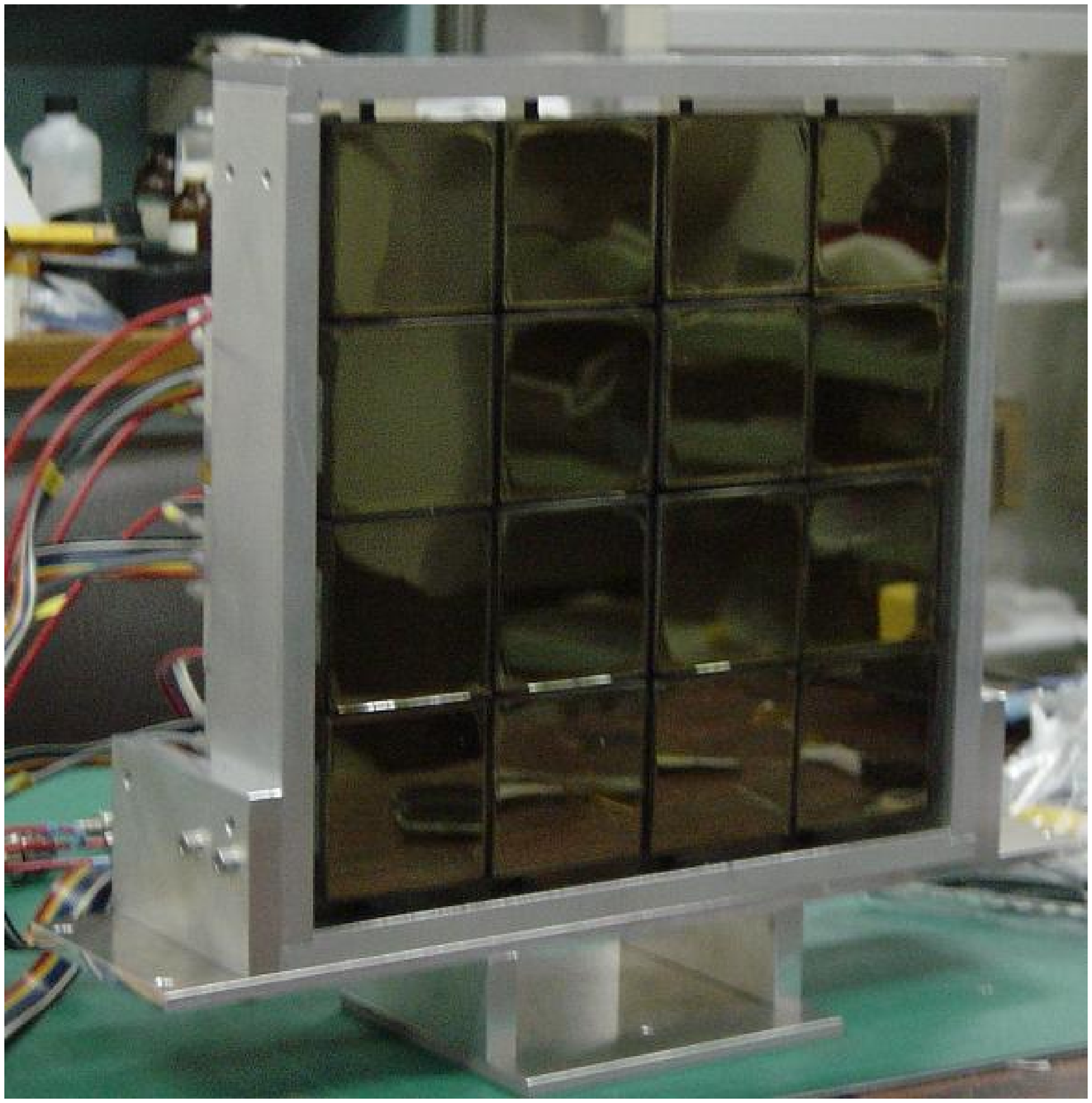}
  \end{center}
  \caption{Photon detector, an array of 16 H8500 PMTs, mounted at 
a 52.5~mm pitch.}
  \label{fig:flat-panel-PMT}
\end{figure}

The photon detector for the tested prototype RICH counter employed $64$ channel 
multi-anode PMTs (Hamamatsu H8500, so called flat-panel PMT) because of their 
large effective area. 16 PMTs were used in a $4 \times 4$ array and aligned with a 
$52.5~{\rm mm}$ pitch, as shown in Figure \ref{fig:flat-panel-PMT}. 
The surface of each PMT is divided into $64$ ($8 \times 8$) channels with a 
$6.0 \times 6.0~{\rm mm}^2$ pixel size. Therefore, the effective area of photon 
detection is increased to $84 \%$. At the back of each PMT, an analog memory 
board is attached to read out multi-channel PMT signals, as described below. Among 
$16$ PMTs, $8$ PMTs were delivered in January, 2002, and the remaining PMTs 
were delivered in October, 2002. Since the manufacture method of the PMT was still 
under development, they exhibit a large variation in quantum efficiency 
and gain. The quantum efficiency at 400~nm varies between $16$\% and $25$\%; 
the gain varies from $1 \cdot 10^6$ to $6 \cdot 10^6$ when the maximal allowed 
high voltage of $-1100$~V is applied to the photo-cathode \cite{Hamamatsu}. 
The PMTs from the later batch show a slightly better performance.

\subsection{Aerogel radiator}

The hydrophobic form of the aerogel radiator from Novosibirsk \cite{Novosibirsk} 
is known to have a long transmission length. However we prefer hydrophobic aerogel 
than hydrophilic one for the application to a collider experiment. 
With a low refractive index ($n=1.01\sim1.03$), such an aerogel was developed 
for the Belle-ACC, and is characterized by a high transmission length 
($\sim 40$~mm at a wave length of 400~nm) which was not achieved before. 
However, the transmission length of aerogel with a higher refractive index of 
$n = 1.05$ fell below one half the value compared to the aerogel with $n = 1.03$. 
Therefore, we reexamined the aerogel production technique in a joint development 
with Matsushita Electric Works Ltd. 
As a result, we found that the important factors determining the transmission length 
are the solvent and selection of the precursor to be used for its production. 
Originally, we used methyl-alcohol for the solvent, and methyl-silicate as a 
precursor \cite{BelleAerogel}. When we applied di-methyl-formamide (DMF) 
\cite{MatsushitaDMF}, and changed the supplier of the precursor, we could improve 
the transmission length of the aerogel. 

\begin{figure}[htbp]
  \begin{center}
    \includegraphics[keepaspectratio=true,width=80mm]{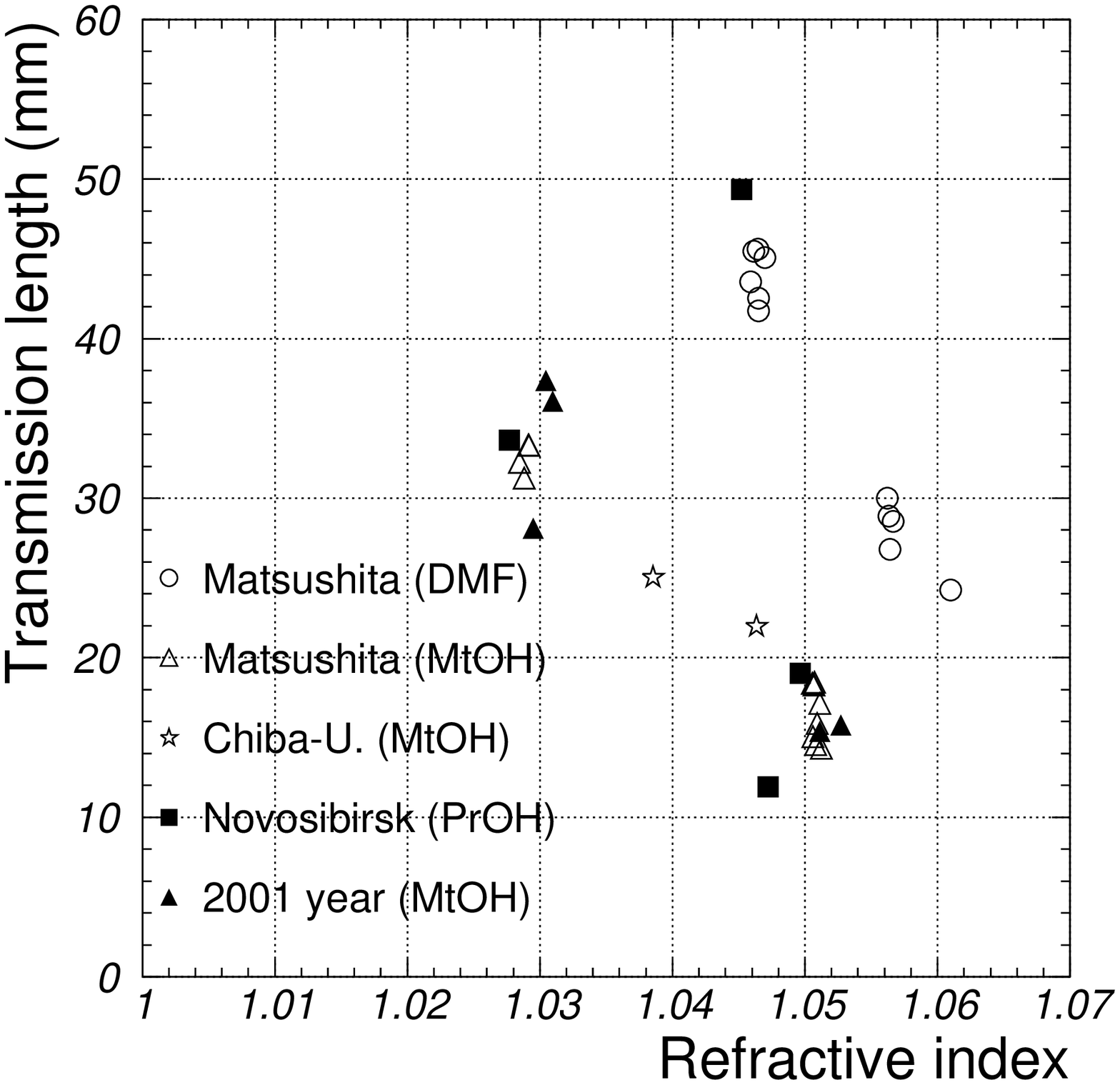}
  \end{center}
  \caption{Transmission length at 400~nm and refractive
index at 405~nm for the aerogel samples used in the test.}
  \label{fig:aero}
\end{figure}

Figure \ref{fig:aero} shows the refractive indices of aerogel and the relation 
to transmission length for samples which were used in this beam test. The 
refractive index was determined by measuring the deflection angle of laser light 
(laser diode: $405~{\rm nm}$) at a corner of each aerogel tile; the transmission length was measured with a photo-spectrometer (Hitachi U-3210). In addition to the 
samples produced with the new technique at Matsushita Electric Works Ltd. and 
Chiba university, samples from BINP (Novosibirsk) were tested \cite{Novosibirsk}; 
for comparison, we also tested the samples used in the previous beam test. 
The thicknesses of the prepared aerogel samples ranged from $10~{\rm mm}$ to 
$25~{\rm mm}$. Various thickness of up to about $50~{\rm mm}$ were tested by stacking 
these samples. Note that in the production of the aerogel samples at BINP propenol 
was used as the solvent, and the resulting aerogel was hydrophilic. Also note that 
the Matsushita aerogel samples produced with the new technique have a very 
similar transmission length as the BINP samples. The transmission length for 
$n{\sim}1.05$ samples used in the first beam test was around $15$~mm, but 
was increased to $45$~mm for Matsushita's sample with the new production method.

\subsection{Beam set-up}

For the beam test, pions with momenta between $0.5~{\rm GeV}/c$ and 
$4~{\rm GeV}/c$ were used. Beside the RICH detector under study, counters for 
triggering, tracking and particle identification were employed.

\begin{figure}[htbp]
  \begin{center}
    \includegraphics[keepaspectratio=true,width=120mm]{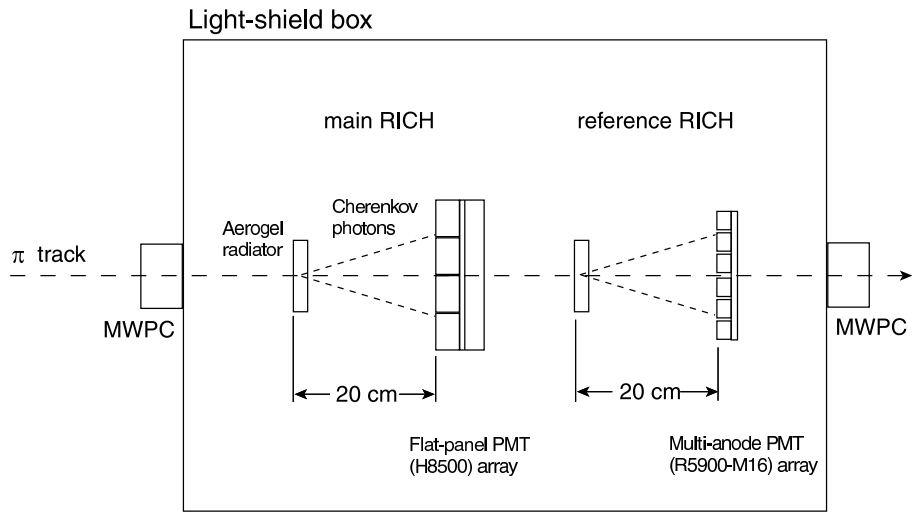}
  \caption{Experimental set-up.}
  \label{fig:rich}
  \end{center}
\end{figure}

The set up of the aerogel RICH is shown in Figure \ref{fig:rich}. Two RICH 
counters were placed in a light-shield box and tested simultaneously. Each RICH 
was composed of a layer of aerogel radiator and a photo-detection plane, parallel 
to the radiator face at a distance of $20~{\rm cm}$. The upstream Cherenkov counter 
was the detector under study; the downstream counter was the one employed in the 
previous beam test. Since the latter uses a well-known photo-detector, multi-anode 
PMTs Hamamatsu R5900, we regarded it as a reference.

Particle identification was done to remove particles other than pions. Two CO$_2$ 
gas Cherenkov counters in the beam line were used to exclude electrons. Also, 
an aerogel counter was equipped and used to exclude protons for the high-momentum 
region. This detector was also used to exclude muons for the low-momentum region 
around $0.5~{\rm GeV}/c$.

The particle trajectories were measured with multi-wire proportional chambers 
(MWPC) at the upstream and downstream ends of the light-shield box. 
These $5 \times 5~{\rm cm}^2$ MWPCs, with $20~\mu$m diameter, gold-plated tungsten 
anode wires at $2~{\rm mm}$ pitch and with $90 \%$ Ar + $10 \%$ ${\rm CH}_4$ gas 
flow, were read out by delay lines on the x and y cathode strips.

The trigger signal was generated as a coincidence of signals from several 
$5 \times 5~{\rm cm}^2$ plastic scintillation counters and anode signals from 
the MWPCs to ensure valid tracking information.

\subsection{Readout electronics}

\begin{figure}[htbp]
 \begin{center}
    \includegraphics[keepaspectratio=true,width=120mm]{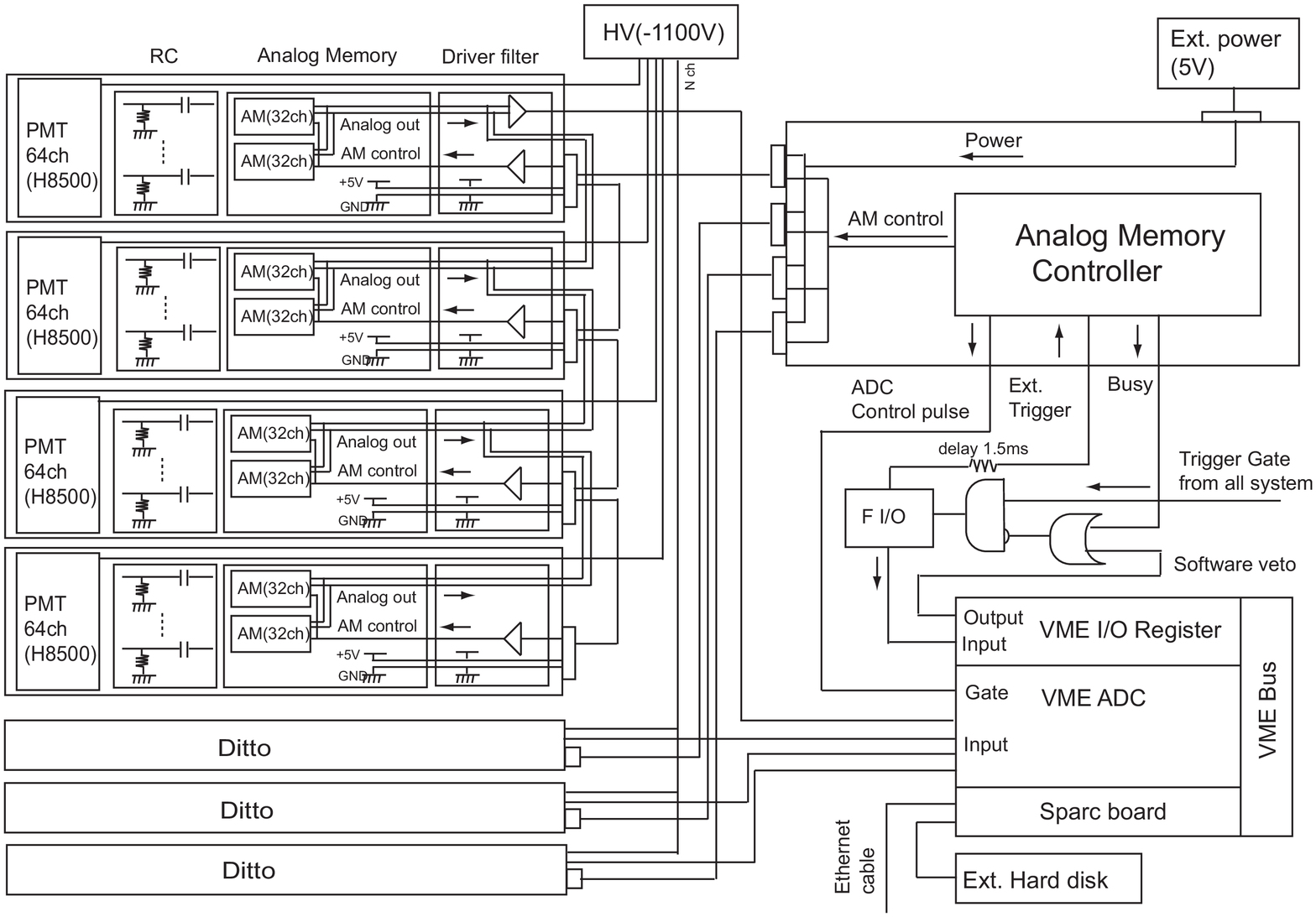}
 \end{center}
  \caption{Schematics of the read-out system for the flat-panel PMTs.}
  \label{fig:rich-readout}
\end{figure}

For the beam test, a new read-out system was designed by using analog memory chips. 
The analog memory chip is based on a chip developed by H.~Ikeda \cite{AnalogMemory} 
for a cosmic-ray experiment. We borrowed the chips from NASDA (National Space  Development Agency of Japan), and developed the chip control system. In the 
analog memory chip, the signals of 32 channels are preamplified, sampled in 
$1~\mu$s intervals, and stored in an 8 steps deep analog pipeline. 
Figure \ref{fig:rich-readout} shows a schematic view of the readout system 
with these analog memories. Two  32 channel analog memories are attached to each 
64 channel PMT. The memories corresponding to four PMTs are controlled by a 
256 channel memory controller. When the gate pulse is formed from the trigger 
signal, a control signal is sent from the controller to the analog memories. 
The difference in the value of the analog memory between the latest and the first 
memory content is fed to the output. The obtained output values of 256 channels 
are clocked into one signal train with a period of 10~$\mu$s per channel. 
Each analog memory controller outputs the serial signal together with 
synchronized control signals. These signals are then read by a 12-bit VME 
ADC (DSP8112, MTT Co.) with a conversion time of 5~$\mu$s.

\subsection{Reference RICH}

A reference RICH was instrumented with multi-anode PMTs, Hamamatsu R5900-M16, the 
same photon detector as used in the previous test \cite{1stBeamTest}. The 
quantum efficiency of the PMTs is around 26\% (at 400~nm), and the gain was around  
$6 \cdot 10^6$ with $-900$~V applied to the photo-cathode. The PMTs were grouped 
in a $2 \times 6$ array at a $30$~mm pitch. Due to a limited number of available PMTs 
and read-out channels, only a part of the Cherenkov ring was covered with photon 
detectors.

\section{Measurement and results}

Most of the test measurements were performed with a $\pi^-$ beam at $3~{\rm GeV}/c$. 
To systematically evaluate the detector performance, data were taken with 
different aerogel samples with various transmission lengths and thicknesses. Data 
were also taken by varying the $\pi^-$ momentum in the range from $0.5~{\rm GeV}/c$ 
to $4.0~{\rm GeV}/c$.    

\begin{figure}[htbp]
\begin{center}
    \includegraphics[keepaspectratio=true,width=120mm]{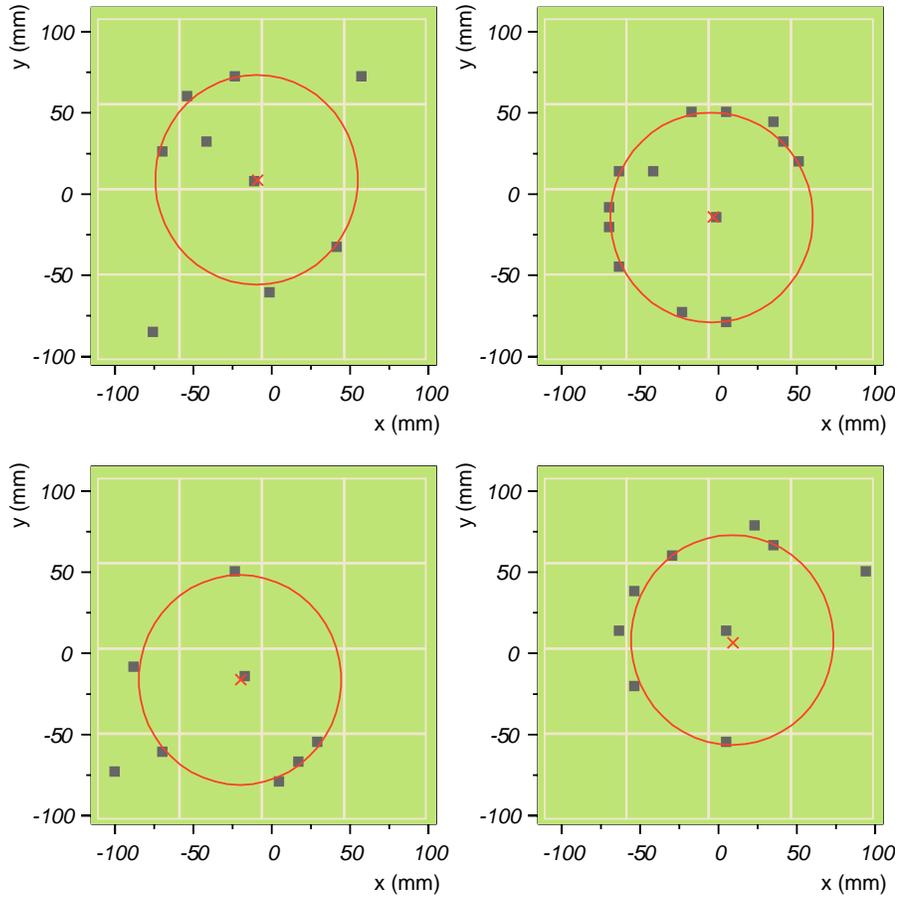}
\end{center}
  \caption{Some examples of event hit patterns for $3~{\rm GeV}/c$ pions. 
  The circle corresponds to the Cherenkov ring as expected from
  the measured beam particle track. The dot corresponds to the impact point of 
  the track upon the PMT window.}
  \label{fig:rich-eventdisplay}
\end{figure}

\begin{figure}[htbp]
\begin{center}
  \includegraphics[keepaspectratio=true,width=80mm]{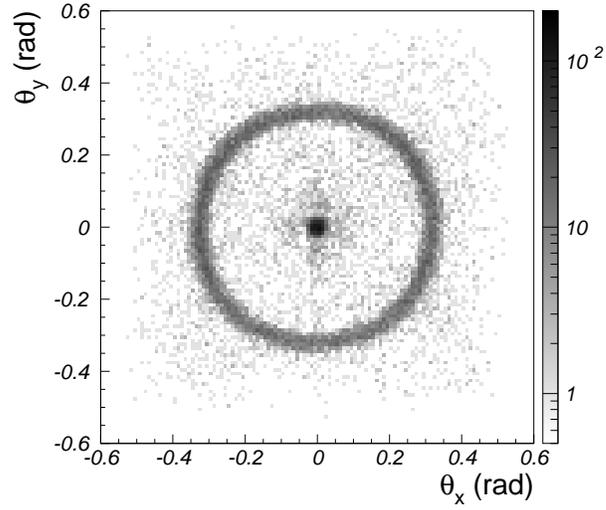}
\end{center}
  \caption{Distribution of PMT hits in the Cherenkov x, y space 
  for $3~{\rm GeV}/c$ pions.}
  \label{fig:typical1}
\end{figure}

A few typical events are displayed in Figure \ref{fig:rich-eventdisplay}. 
The hits on PMTs can be associated with the expected position of the Cherenkov ring. 
The hit near the center of the ring is due to Cherenkov radiation generated 
by the beam particle in the PMT window. The distribution of accumulated hits is shown 
in Figure \ref{fig:typical1}. Cherenkov photons from the aerogel radiator are 
clearly seen with a low background level. The background hit distribution on the 
photon detector is consistent with the assumption that it originates from 
Cherenkov photons which were Rayleigh scattered in the radiator.

\begin{figure}[htbp]
\begin{center}
   \includegraphics[keepaspectratio=true,width=80mm]{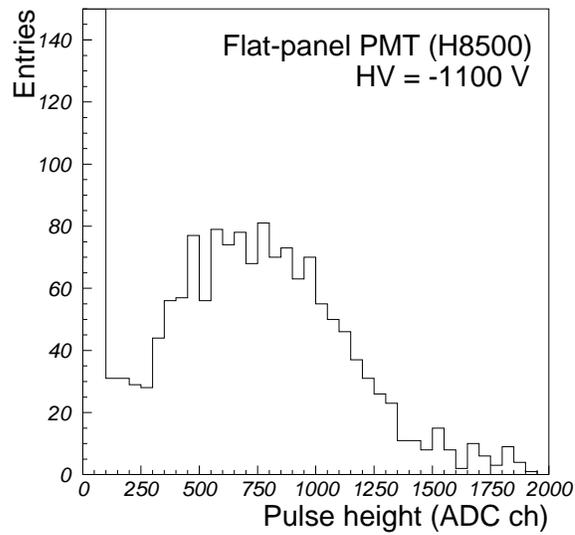}
\end{center}
  \caption{Pulse-height distribution for the flat-panel PMT (H8500) 
for the hits in the region within $3\sigma$ of the mean Cherenkov angle. 
Data were corrected with the procedure described in the text. 
In this figure, the pulse-height distribution for the high sensitive PMT is shown.
For further analysis, we used the hits above a threshold ADC value, 120. 
}
  \label{fig:adc}
\end{figure}

The pulse-height distribution of the Cherenkov photons detected in one of the 
flat-panel PMT is shown in Figure \ref{fig:adc}. The raw data were corrected 
as follows. A common-mode fluctuation of the base line was subtracted and signals 
due to cross-talk in the read-out electronics were removed. The signal mainly 
containing one photoelectron is clearly separated from the pedestal peak. Note, 
however, that this distribution differs considerably from tube to tube because of the large variation in performance, as described before. For further analysis we 
also applied a threshold cut to suppress the pedestal noise contribution. 

\subsection{Cherenkov-angle resolution for single photons}

\begin{figure}[htbp]
\begin{center}
    \includegraphics[keepaspectratio=true,width=120mm]{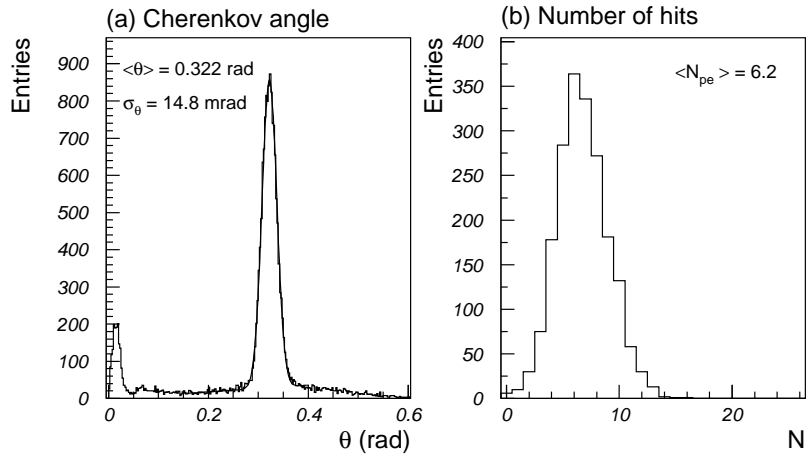}
\end{center}
   \caption{Distribution over the Cherenkov angle for single photons (a), 
   and the number of detected photons per ring (b), for a $20$~mm thick 
aerogel radiator sample with $n=1.056$ and a transmission length of $30$~mm.}
 \label{fig:typical2}
\end{figure}

\begin{figure}[htbp]
\begin{center}
    \includegraphics[keepaspectratio=true,width=80mm]{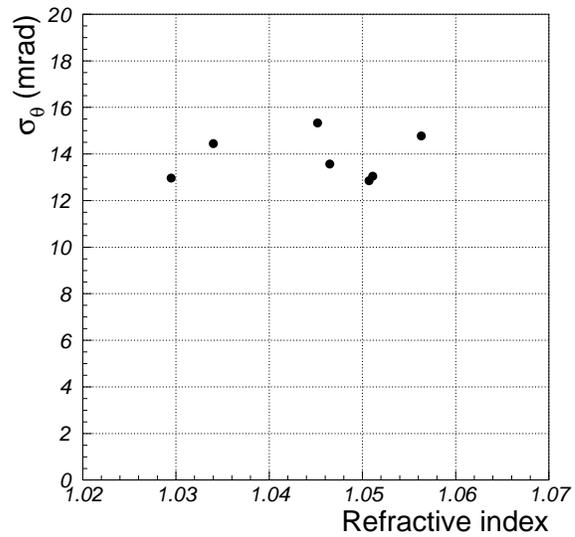}
\end{center}
  \caption{ Typical single photon resolution for $20$~mm thick aerogel samples.}
  \label{fig:s2n}
\end{figure}

\begin{figure}[htbp]
\begin{center}
    \includegraphics[keepaspectratio=true,width=80mm]{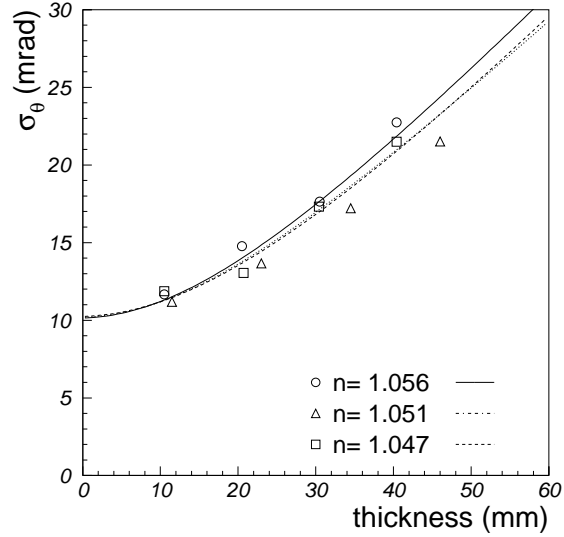}
\end{center}
  \caption{Angular resolution as a function of the aerogel thickness.
The symbols correspond to the data for the different samples and the 
curves are function described in the text (also including the 
unknown contribution of about 6 mrad).}
  \label{fig:s2d}
\end{figure}

\begin{figure}[htbp]
\begin{center}
    \includegraphics[keepaspectratio=true,width=80mm]{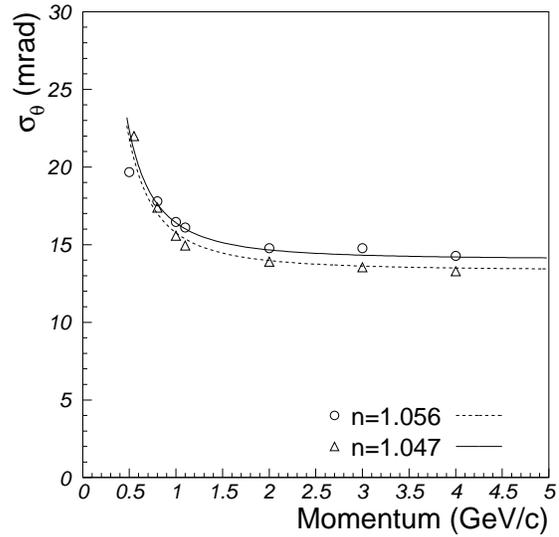}
\end{center}
  \caption{Angular resolution as a function of the charged particle 
  momentum for two different samples. The symbols correspond to the data 
  and the curves are fits including the effect of multiple-scattering.}
  \label{fig:s2p}
\end{figure}

Figure \ref{fig:typical2}(a) shows a typical distribution of the Cherenkov-angle 
for single photons. The angular resolution was obtained from a fit of this 
distribution with a Gaussian signal and a linear function for the background. 
Figure \ref{fig:s2n} shows the resolution in the Cherenkov angle for the 
$\pi^-$ beam at $3~{\rm GeV}/c$ and $20$~mm thick aerogel samples. The resolution 
was around $14$~mrad, independent of the refractive index. The main contributions to 
the resolution of the Cherenkov angle come from the uncertainty in the emission point 
and from the pixel size of the PMT. The first contribution is estimated to be  
$\sigma_{emp} = d \sin{\theta_c}\cos{\theta_c}/L\sqrt{12}$, where $d$ is 
the aerogel thickness, $\theta_c$ is the Cherenkov angle and  $L$ is the distance 
from an average emission point in the aerogel to the surface of the PMT. The 
second contribution is $\sigma_{pix} = a \cos^2{\theta_c}/L\sqrt{12}$, where $a$ 
is the pixel size. The measured variation of the resolution with the thickness 
of aerogel is shown in Figure \ref{fig:s2d}. 
By comparing the measured resolution and the expected values, we observed a 
rather good agreement. There was, however, a discrepancy between the two, which can 
be accounted for by a contribution of about $6$~mrad. The discrepancy could arise 
from the effect of aerogel (non-flat aerogel surface and possible non-uniformities 
in the refractive index due to position variation and chromatic dispersion), which 
are subject to further investigation. The uncertainty in the track direction 
is expected to be negligible at $3~{\rm GeV}/c$, but increases considerably at 
low momenta ($0.5~{\rm GeV}/c$) due to the effect of multiple-scattering, as can 
be seen in Figure \ref{fig:s2p}.

\subsection{Photoelectron yield}

\begin{figure}[htbp]
 \begin{center}
    \includegraphics[keepaspectratio=true,width=80mm]{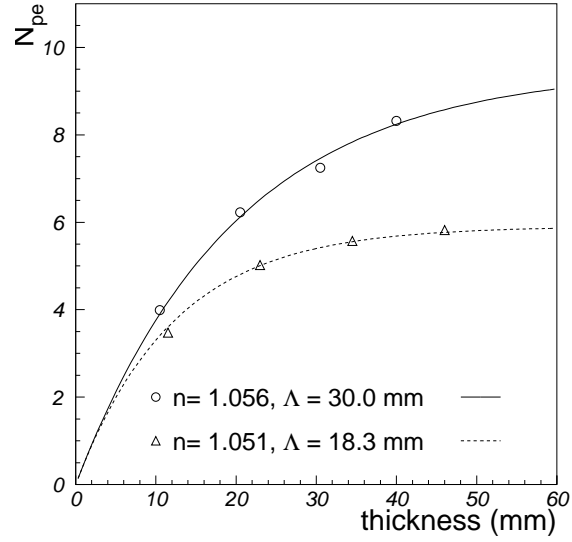}
 \end{center}
  \caption{Number of detected photons per Cherenkov ring 
  as a function of the aerogel thickness. The symbols correspond 
  to the data and the curves are fits described in the text.}
  \label{fig:n2d}
\end{figure}

\begin{figure}[t]
 \begin{center}
    \includegraphics[keepaspectratio=true,width=80mm]{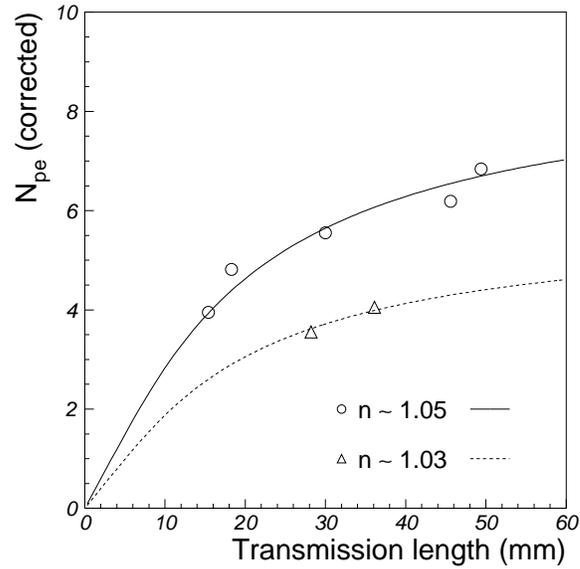}
 \end{center}
  \caption{Number of detected photons per Cherenkov ring ($N_{pe}$) for $20$~mm thick 
  aerogel samples as a function of the transmission length. $N_{pe}$ is corrected 
  for the refractive index to $n = 1.05$ and $n = 1.03$ respectively.  
  The symbols correspond to the data and the curves are fits described in the text.}
  \label{fig:n2tr}
\end{figure}

\begin{figure}[t]
  \begin{center}
    \includegraphics[keepaspectratio=true,height=70mm]{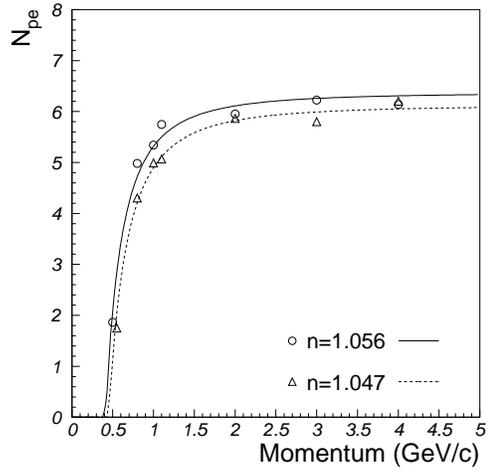}
  \end{center}
  \caption{Number of detected photons per Cherenkov ring as a function of 
  the charged particle momentum. The symbols correspond to the data and the 
  curves are fits described in the text. }
  \label{fig:n2p}
\end{figure}

Figure \ref{fig:typical2}(b) shows a typical distribution of the number of 
hits  within $\pm 3\sigma$ from the average Cherenkov angle. The number of hits for 
the signal region was estimated by subtracting the background from the fits to 
the Cherenkov-angle distribution. The number of detected photons ($N_{pe}$) depends 
on the aerogel thickness and the effect of scattering. It is expressed as \\
$N_{pe} = C \int_0^d \int \epsilon(\lambda)\lambda^{-2} \sin^2\theta_c \exp({-\frac{x}{\Lambda(\lambda)\cos\theta_c}})d\lambda dx
\approx C^{\prime} \sin^2\theta_c \Lambda \cos\theta_c( 1 - \exp({-\frac{d}{\Lambda \cos\theta_c}}))$, 
where $\Lambda$ is the transmission length of the aerogel at an average wave 
length of $400$~nm and $\epsilon(\lambda)$ is quantum efficiency of the PMT. 
Figure \ref{fig:n2d} shows the dependence of the number of detected photons 
on the aerogel thickness. As expected from the above expression, the number of 
photons does not linearly increase with the aerogel thickness, but saturates due 
to the scattering effect in aerogel. Figure \ref{fig:n2tr} shows the 
dependence of the number of photons with transmission length. From the figure 
the benefit of the improvement in the transmission length of the $n=1.05$ aerogel 
from around $15$~mm, as used in the previous beam test, to $45$~mm using the 
new production technique becomes evident. The dependence on the pion 
momentum, displayed in Figure \ref{fig:n2p}, is fitted with the form 
expected from the Cherenkov relations, and shows a good agreement. For pions 
with momenta above $1~{\rm GeV}/c$, the number of detected Cherenkov photons 
was typically around 6 for aerogel samples with $n=1.05$.

The performance of the RICH counter under study was compared in the same set-up 
with the performance of the reference counter with a well-known photon detector, 
Hamamatsu R5900-M16 multi-anode PMTs. Since the two counters have a different 
active area fraction ($84$\% for the flat-panel PMTs, and $36$\% for the 
R5900-M16 PMTs) and a different acceptance, the comparison of the photon yields 
was made by normalizing to the full active surface. While the flat-panel yield for 
a particular case was $6.2$, which resulted in $7.8$ if extrapolated to the full active area, the corresponding number for the R5900-M16 was $12$. It appears that 
this difference is mainly due to the rather low quantum efficiency and 
amplification of some of the flat-panel tubes employed. This, in turn, 
causes inefficiencies in single photon detection with a given threshold. If the 
best tube in the set is normalized to the full acceptance, the corresponding 
number increases to $10$, and we would expect about 8 photons per ring.

\subsection{Particle Identification} 

\begin{figure}[t]
  \begin{center}
    \includegraphics[keepaspectratio=true,width=80mm]{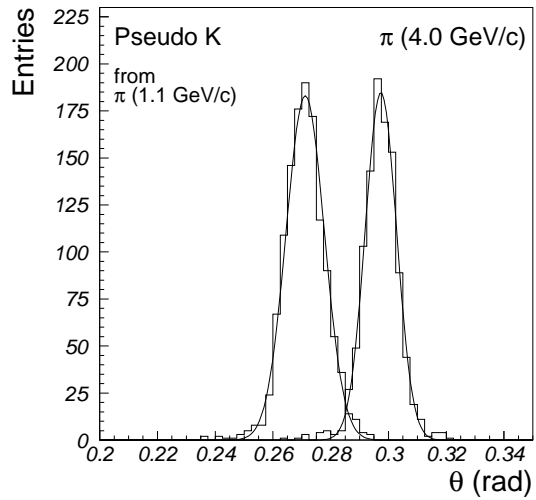}
  \end{center}
  \caption{Cherenkov angle per track for pions of $4.0~{\rm GeV}/c$ and 
  $1.1~{\rm GeV}/c$. Pions at $1.1~{\rm GeV}/c$ are used to represent the 
  kaon beam of $4~{\rm GeV}/c$. The angular resolutions for $4.0~{\rm GeV}/c$ 
  and $1.1~{\rm GeV}/c$ are $5.4$~mrad and $6.7$~mrad and two peaks are 
  separated by $4.2\sigma$. 
  }
  \label{fig:sep_pik_4gev}
\end{figure}

Finally, we estimate the performance of pion/kaon separation in the momentum 
range of around $4~{\rm GeV}/c$, which is of considerable importance for the 
Belle experiment. If we take into account a typical measured value for the 
single-photon angular resolution, $\sigma_c \sim 14$~mrad, and the number of 
detected photons $N_{pe} \sim 6$, typical for 20~mm thick aerogel samples with 
$n = 1.05$, we can estimate the Cherenkov angle resolution per track to be 
$\sigma_c/\sqrt{N_{pe}} = ~5.7$~mrad. This naive estimate is also confirmed by 
the direct measurement shown in Figure \ref{fig:sep_pik_4gev}. 
Here, the track-by-track Cherenkov angle is calculated by taking the average of 
the angles measured for hits around the predicted position of the Cherenkov 
ring. From this we can deduce that at $4~{\rm GeV}/c$, where the difference 
of Cherenkov angles for pions and kaons is $23$~mrad, a $4\sigma$ separation 
between the two is possible. As an additional cross check, we have also collected 
data with the pion beam of $1.1~{\rm GeV}/c$, which can be used to represent a 
kaon beam of $4~{\rm GeV}/c$ (apart from a slightly larger sigma due to 
multiple scattering). As can be seen from Figure \ref{fig:sep_pik_4gev}, the 
two peaks are well separated. Thus, the proximity focusing aerogel RICH seems to 
be promising for the upgrade of the Belle PID system at the forward region.

\section{Conclusions}

We report on the test beam results of a proximity focusing RICH using aerogel as 
the radiator. To obtain larger photoelectron yields, we used flat-panel multi-anode 
PMT with a large effective area, and aerogel samples produced with a recently 
developed method which have a higher transmission length than before. We also 
developed a multi-channel PMT read-out system using analog memory chips. A 
clear Cherenkov ring from the aerogel radiator could be observed, and the number 
of photons was enhanced compared to that in previous tests. We performed a 
systematic study of the detector using various samples of the aerogel. The 
typical angular resolution was around $14$~mrad and the number of detected 
photoelectrons was around $6$. The pion/kaon separation at $4~{\rm GeV}/c$ is 
expected to be around $4\sigma$.

However, we still have some issues which have to be solved for implementation in 
the Belle spectrometer. The most important item is the development of a PMT which 
can be operated under a strong magnetic field (1.5T). An example of a candidate for 
such a device is a multi-anode hybrid photodiode (HPD) or hybrid avalanche 
photodiode (HAPD). Of course, for a good candidate, its ability 
to efficiently detect single photons on a large active area has to be demonstrated. 
The other item is mass production of the aerogel tiles. While we have demonstrated 
that the new production method significantly increases the transmission length of the 
$n = 1.05$ aerogel, the production method has to be adapted to stable 
good-quality manufacturing. We will study these items at the next stage towards construction of a real detector.

\section{Acknowledgment}

We would like to thank the KEK-PS members for operation of accelerator and 
for providing the beam to the $\pi2$ beam line. 
We also thank H.~Ikeda (KEK) and the Meisei Co. 
for their help in preparing the read-out electronics, 
the Matsushita Electric Works Ltd. for the good collaboration in developing 
the new aerogel type, and Hamamatsu Photonics K.K. for their support in equipping 
the photon detector.
We also thank A.Bondar (BINP, Novosibirsk) for providing us excellent 
aerogel samples, and Dr. T.Goka of NASDA for providing us their read-out chips.
One of the authors (T.M.) is grateful to Fellowships of the Japan 
Society for the Promotion of Science (JSPS) for Young Scientists. 
This work was supported in part by a Grand-in-Aid for Scientific
Research from JSPS and the Ministry of Education, Culture, Sports, Science and 
Technology under Grant No. 13640306, 14046223, 14046224, and in part by 
the Ministry of Education, Science and Sports of the Republic of Slovenia.

\end{document}